%% file: spherical.tex
\documentclass[twoside]{amsart}
\usepackage{url}
\usepackage[latin1]{inputenc}
\usepackage{graphicx}
\usepackage{color}
\usepackage{pstricks}
\def\cM{{\mathcal M}}
\def\cL{{\mathcal{L}}}
\def\RR{{\mathbb R}}
\def\thetab{\boldsymbol{\theta}}
\def\omegab{\boldsymbol{\omega}}
\def\Sigmab{\boldsymbol{\Sigma}}
\def\alphab{\boldsymbol{\alpha}}
\def\betab{\boldsymbol{\beta}}
\def\gammab{\boldsymbol{\gamma}}
\def\varpib{\boldsymbol{\varpi}}
\def\varthetab{\boldsymbol{\vartheta}}
\def\Omegab{\boldsymbol{\Omega}}
\def\dd{\mathbf{d}}

\author{Ronny Richter}
\email{richter@tat.physik.uni-tuebingen.de}
\address{Institut für Astronomie und Astrophysik, Universität
  Tübingen, Auf der Morgenstelle 10, 72076 Tübingen, Germany}
\author{Jörg Frauendiener}
\email{joerg.frauendiener@uni-tuebingen.de}
\address{Institut für Astronomie und Astrophysik, Universität
  Tübingen, Auf der Morgenstelle 10, 72076 Tübingen, Germany}
\author{Marlene Vogel}
\curraddr{Institut für Atomare Physik und Fachdidaktik,
  Technische Universität Berlin, 
  Hardenbergstraße 36, 10623 Berlin, Germany}
\email{mvogel@physik.tu-berlin.de}
\address{Institut für Astronomie und Astrophysik, Universität
  Tübingen, Auf der Morgenstelle 10, 72076 Tübingen, Germany}

\begin{document}
\title[Discrete Differential Forms in General Relativity]{Application of
  Discrete Differential Forms to Spherically Symmetric 
  Systems in General Relativity}

\begin{abstract}
  In this article we describe applications of Discrete Differential Forms
  in computational GR. In particular we consider the initial value problem in
  vacuum space-times that are spherically symmetric. The motivation to
  investigate this method is mainly its manifest coordinate independence.
  Three numerical schemes are introduced, the results of which are
  compared with the corresponding analytic solutions. The error of two
  schemes converges quadratically to zero. For one scheme the errors
  depend strongly on the initial data.
\end{abstract}
\maketitle

\section{Introduction}
\label{sec:intro}
\thispagestyle{empty}

Most methods that are presently used in numerical GR are in some sense
referred to a coordinate system. This can be a major
problem, because not only is it impossible in general to cover a
global space-time with a single coordinate chart. But also it is
generally impossible to know beforehand the effects that certain
gauge conditions specified during the course of a simulation will imply.

In view of this the question occurs, as to whether it is possible to
develop a numerical method that is manifestly coordinate invariant.
One such method is Regge calculus~\cite{regge1961:_gener_relat} which,
unfortunately, so far has not played a role in computational GR (see
however~\cite{gentle2002:_regge}). In other approaches to treat the problem
of coordinate dependencies multiple coordinate systems
are used to cover the space-time
\cite{Thornburg:cqg21_3665,Caltech:gr-qc/0607056,LSU:cqg21_S553}.

However, a manifestly invariant numerical method
must be based on invariant quantities describing the geometry of
space-time. The prime examples for invariant quantities on a manifold
are the scalar fields, but in the usual description even they are
coordinate dependent, because the description of the points of the
manifold themselves depends on the choice of a coordinate system.
Therefore, in order to avoid coordinates we must not even use
coordinates for the localisation of points of the space-time manifold.
This implies that we cannot use the usual definition of a manifold as
a collection of coordinate charts with transition functions which is
the basis of almost all analytical and numerical treatments of the
Einstein equations.

In the usual procedure for discretisation the manifold structure is
untouched while the equations are discretised, i.e., evaluated only for
a finite number of points of the manifold. When the use of coordinates
is to be avoided one has to start the discretisation at an even lower
level namely on that of the manifold itself. Hence, on a quite basic
level the space-time can be considered as a collection of abstract
objects called points. The structures on the space-time are then
described as certain relations between these points. In our case the
relevant structure consists of primarily topological and geometric
relationships. For the present purpose we find it more reasonable to
consider the topological relationships as given in advance so that the
aim of computational GR is to find the geometric
relations between the points based on an appropriate formulation of
the Einstein equations.

To achieve this we continue the work presented in
\cite{frauendiener2006:_discr_differ_forms_gener_relat}.
That is we approximate a manifold and its differentiable structure by a
cellular paving \cite{bossavit1998:_discr_em_prob},
i.e. a collection of finitely many
cells. The cells are the images of a certain number of standard shapes
like (hyper-)cubes or
$n$-simplices. In the case where all standard shapes are simplices
we talk about a triangulation and the cellular paving is a simplicial
complex. The cellular paving is supposed to
have the same topological properties as the envisaged space-time.

To illustrate the idea we consider the example of a standard 3-simplex
which can be viewed as the interior of a tetrahedron. It is a
3-dimensional manifold. Its boundary is composed of four 2-simplices
(faces), six 1-simplices (edges) and four 0-simplices (nodes). With
such $p$-simplices we can associate several quantities which can be
interpreted in a physical way. Examples are the charge inside a
volume, a flux through a face, the work done along an edge or the
value of a potential at a given point. In all these cases we associate
\emph{numbers} with a simplex and these numbers are usually obtained
by \emph{integration}, i.e., by adding up contributions from
`infinitesimal' pieces making up the finite simplex. So, in each case
we obtain a map from $p$-simplices to numbers.

Differential $p$-forms can be viewed as `the objects which are
integrated over $p$-dimensional submanifolds' so they provide maps
from $p$-dimensional submanifolds to the reals. Thus, the maps
presented above correspond to differential forms, but restricted to
$p$-simplices. These objects are known as discrete differential forms.
They have received some attention since
Bossavit~\cite{bossavit1988:_mixed_whitn} had pointed out that they
correspond to the lowest order mixed finite element spaces defined by
N\'ed\'elec~\cite{nedelec1980:_mixed_r} (see
also~\cite{raviartthomas1977:_mixed_fem}). Finite elements of mixed
type have been used successfully in numerical applications to
electrodynamics, see~\cite{bossavit1998:_discr_em_prob,%
hiptmair2002:_finit_elem_disc_em, bossavit1998:_comput_elect}. In
numerical GR the finite element method is used e.g. in
\cite{PSU:PRD73_044028}.

Our task is now to relate geometric properties such as lengths,
angles, holonomies and curvature using differential forms to the
triangulation and the various parts of the simplices respectively. Since
$0$-forms are functions they describe properties at single points. In
order to formulate relations between points such as the distance
between two points or the holonomy around a loop we need $p$-forms with
$p>0$.

In order to use this approach one needs to have a formulation of
geometries and, in particular, of GR which uses differential forms.  A
formulation of geometries based on differential forms has been
provided by
\'E.~Cartan~\cite{cartan2001:_rieman_geomet_in_orthog_frame}. The
further step towards a formulation of GR using differential forms has
been carried out by several authors. We mention here the work of
Sparling~\cite{sparling2001:_twist_einst} who has set up an exterior
differential system of equations which is closed if and only if the
vacuum Einstein equations hold.
In~\cite{frauendiener2006:_discr_differ_forms_gener_relat} we have
shown in detail how to set up the discrete formalism based on this
exterior system using the ideas explained above.

In summary, the variables of our proposed discrete formulation will be
the integrals of differential forms over submanifolds. In order to get
a finite number of variables we use a finite number of these
submanifolds based on a triangulation of the computational domain and
discretise a description of GR that uses (finitely many) differential
forms.
 
The formulation of GR that we use is based on the Cartan formalism of
moving frames and Sparling's exterior system for vacuum GR. In this
article we describe a simplification of the general formalism which
occurs in spherical symmetry. The plan of the paper is as follows. In
sect.~\ref{spherical_Sparling_equations} we describe the equations
which result from a symmetry reduction. In section~\ref{discrete_form_section}
we present three possibilities to implement these equations in a fully
discrete evolution scheme. In section~\ref{scenarios} we discuss how
the method can be tested and in section \ref{results} we present the
results of those tests. Some final remarks can be found in section
\ref{discussion}.

\section{The spherically symmetric equations}
\label{spherical_Sparling_equations}

We start with the formulation of GR using exterior
forms~\cite{sparling2001:_twist_einst}.  The basic variables in this
formalism are the four 1-forms of a pseudo-orthonormal tetrad
$\thetab^i$, $i=0,\ldots,3$ \cite{Wikipedia:Tetrad}. Together with the
coefficients $\eta_{ik}=\text{diag}(1,-1,-1,-1)$ they define the
metric as 
\begin{equation}
g=\thetab^0\otimes\thetab^0-
\thetab^1\otimes\thetab^1-
\thetab^2\otimes\thetab^2-
\thetab^3\otimes\thetab^3=
\eta_{ik}\thetab^i\otimes\thetab^k.
\end{equation}
For the description of the connection in this formalism sixteen
1-forms ${\omegab^i}_k$, $i,k=0,\ldots,3$ are used. The
connection should be compatible with the metric and torsion free,
which translates into the antisymmetry requirement and the first
Cartan equation, respectively\footnote{Here and in what follows it is
understood, that the product of differential forms is the
anti-symmetrised tensor product, i.e. the exterior product.}:
\begin{align}
\label{eq:torsion-free}
\eta_{ik}{\omegab^k}_j+\eta_{jk}{\omegab^k}_i&=0,&
\dd\thetab^i+{\omegab^i}_k\thetab^k&=0.
\end{align}
Furthermore the metric should fulfil Einstein's field equations, which
is equivalent to
\begin{align}
\label{eq:Sparling}
\mathbf{dL}_i=\mathbf{S}_i+8\pi{T_i}^k\Sigmab_k.
\end{align}
Here, $T_{ik}$ is the usual energy momentum tensor and
\[
\Sigmab_i=\frac16\varepsilon_{ijkl}\thetab^j
\thetab^k\thetab^l,
\] 
are the so called hypersurface~3-forms. The forms $\mathbf{L}_i$ and
$\mathbf{S}_i$ are the Nester-Witten 2-form and the Sparling~3-form,
defined by
(see~\cite{frauendiener2006:_discr_differ_forms_gener_relat,sparling2001:_twist_einst}):
\begin{align}
\mathbf{L}_i&=
\frac{1}{2}\varepsilon_{ijkl}{\omegab^{jk}}\thetab^l,&
\mathbf{S}_i&=\frac{1}{2}\varepsilon_{ijkl}\left(\omegab^{jk}
{\omegab^l}_m\thetab^m -
{\omegab^j}_m\omegab^{mk}
\thetab^l\right).
\end{align}
In vacuum, when ${T_i}^{k}=0$, these equations determine the geometry
of space-time. If there is matter, additional matter equations are
needed. However, we will be concerned only with the vacuum case so
that we will have to solve the equations
\begin{subequations}
\label{eq:vacuum_equations}
\begin{align}
\label{eq:vacuum_equations_torsion}
\dd\thetab^i+{\omegab^i}_k\thetab^k&=0,\\
\label{eq:vacuum_equations_Einstein}
\mathbf{dL}_i&=\mathbf{S}_i.  
\end{align}
\end{subequations}
Although the geometry is fixed, there is still the freedom of choosing
a gauge, i.e. there are Lorentz transformations $\Lambda^i{}_k$ of the
tetrad that do not change the metric
\begin{equation}
\label{eq:gauge-freedom}
g=\eta_{ik}\thetab^i \otimes\thetab^k=
\eta_{ik}{\Lambda^i}_j \thetab^j \otimes{\Lambda^k}_l \thetab^l
= (\eta_{ik}\Lambda^i{}_j\Lambda^k{}_l)\,\thetab^j\otimes\thetab^l.
\end{equation}
That means by using the tetrad for the description of geometries we
introduced unphysical (gauge) degrees of freedom. The same problem occurs
when coordinate systems are used.
However, we believe that the tetrad, beeing a geometric object, has a more
intuitive meaning than mere coordinates. Therefore, it might be easier
to choose a useful tetrad than an appropriate coordinate system.

In this work we will concentrate on general relativistic systems with
spherical symmetry.
Thus, we will assume that the rotation group $SO(3)$ acts isometrically on the
space-time and that the orbits of this action are
2-dimensional space-like submanifolds. These are necessarily spheres
whose area we write as $4\pi R^2$.
In appendices \ref{app:spher_symm} and \ref{app:reduction} it
is shown, how to `factor out' the symmetry action i.e., the angular
dependence and how to derive an exterior system on the 2-dimensional
`orbit space' $\cM_1$ spanned by the radial and the time directions.%
\footnote{Only for $R>0$ the orbits are 2-dimensional, which implies
that the symmetry action can not be `factored out' when $R=0$.}

This can be done by a decomposition of the 4-dimensional space-time
manifold into the 2-dimensional spheres and the 2-dimensional
orbit space, followed by some simplifications and results in the following
system
\begin{subequations}
\label{eq:0formsystem}
\begin{align}
\label{eq:0formsystema}
0&= \mathbf df_0 - f_1\omegab
-\frac12 \left( f_1^2 - 3f_0^2 - \frac1{R^2} \right) \thetab^0
+ (f_0f_1) \,\thetab^1,\\ 
\label{eq:0formsystemb}
0&= \mathbf df_1 -
f_0\omegab - \frac12\left(f_0^2 - 3 f_1^2 + \frac1{R^2} \right)
\thetab^1  + (f_0f_1)\,\thetab^0,\\
\label{eq:0formsystemc}
 0&= \mathbf
d\left(\sqrt{R^2}\left(R^2f_0^2-R^2f_1^2+1\right)\right),\\ 
\label{eq:0formsystemd}
0&=
f_0\left( \mathbf
d\thetab^1+{\omegab}\thetab^0\right) -
f_1\left(\mathbf
d\thetab^0+{\omegab}\thetab^1\right).
\end{align}
\end{subequations}
Here $(\thetab^0, \thetab^1)$ is a dyad in the 2-dimensional
orbit space $\cM_1$ which carries a Loren\-tzian metric. The $SO(1,1)$
connection on this space is given by the 1-form
${\omegab}$. It is a consequence of the equations above that this
connection is torsion free. The geometric properties of
the orbits are described by the functions $f_0$, $f_1$ and $R^2$
(e.g. $4\pi R^2$ is the area of the orbit).
For details see appendix \ref{app:spher_symm}.

By introducing the 1-forms
\begin{equation}
\begin{aligned}
 \tilde{\thetab}^0&:=\frac1R \thetab^0,&
\tilde{\thetab}^1&:=\frac1R \thetab^1,\\ 
{\alphab}&:=f_0\thetab^0 + f_1\thetab^1,&
{\betab}&:=f_1\thetab^0 + f_0\thetab^1
\end{aligned}
\end{equation}
the equations (\ref{eq:0formsystem}) can be rewritten as follows
\begin{subequations}
\label{eq:1formsystem}
\begin{align}
  \dd\tilde{\thetab}^0
  +{\omegab}\tilde{\thetab}^1+\alphab\tilde{\thetab}^0&=0,\\
    \dd\tilde{\thetab}^1
  +{\omegab}\tilde{\thetab}^0+\alphab\tilde{\thetab}^1&=0,\\
  \dd\alphab&=0,\\
  \dd\betab+2\alphab\betab+
  \tilde{\thetab}^0\tilde{\thetab}^1&=0,\\
  \dd{\omegab}-\alphab\betab-
  \tilde{\thetab}^0\tilde{\thetab}^1&=0,
\end{align}
\end{subequations}
together with the algebraic relations
\begin{equation}
\label{eq:algebr}
 \alphab\tilde{\thetab}^0
+\betab\tilde{\thetab}^1=0,\qquad
\alphab\tilde{\thetab}^1
+\betab\tilde{\thetab}^0=0.
\end{equation}
The two equations~(\ref{eq:algebr}) are needed to ensure that
$\alphab$ and $\betab$ are constructed out of only two
functions.  They can be interpreted as
\begin{equation}
\label{eq:HodgeInterpretation}
\star\alphab=\betab
\end{equation}
where $\star$ is the 2-dimensional Hodge operator~\cite{frankel1997:_physic}.
In the calculations we will use discrete versions of
both~(\ref{eq:algebr}) and~(\ref{eq:HodgeInterpretation}).
  
From the definition of the Hodge operator it is clear, that
$\star\thetab^1=\thetab^0$. This means that in
(\ref{eq:1formsystem}) we have actually only three 1-forms
$\alphab$, $\thetab^1$ and ${\omegab}$
and their Hodge duals. From now on the Hodge duals will be called dual
forms, whereas the forms $\alphab$, $\thetab^1$ and $\omegab$
themselves get the generic name direct forms. We also have for
every form and its Hodge dual an equation, where its exterior
derivative is involved except for the dual of
${\omegab}$. It turns out, that the exterior
derivative of this form is pure gauge because it can be set to any
value by an appropriate choice of gauge in~$\cM_1$. This property
can be used to fix the gauge.

If the geometry is regular at the origin ($R=0$) then we may also include
it into the computational domain. Thus, the origin becomes a boundary.
We choose the gauge such that the dyad at the origin can be defined as
the limit of $(\boldsymbol\theta^0,\boldsymbol\theta^1)$, i.e. such that
this limit exists.

Then it is easy to verify that if we choose at
$R=0$ an arbitrary vector
$\mathbf V=V^0\mathbf e_0+V^1\mathbf e_1$ with finite components $(V_0,V_1)$,
then $\mathbf{d}R^2(\mathbf V)=0$ and
$\mathbf{d}R(\mathbf V)=V_R<\infty$. It follows from \eqref{eq:2} that
in the limit $R\rightarrow 0$ the expression
$R(f_0\boldsymbol\theta^0+f_1\boldsymbol\theta^1)(\mathbf V)$
must be finite.
It is clear that $R\boldsymbol\theta^0(\mathbf V)= 0$ and
$R\boldsymbol\theta^1(\mathbf V)= 0$ at $R=0$. Thus, the limits of
$Rf_0$ and $Rf_1$ must be finite there, and hence $f_0$ and $f_1$ diverge.
That means we must use the variables $\tilde f_0:=Rf_0$
and $\tilde f_1:=Rf_1$ instead of $f_0$ and $f_1$.

Now we insert this observation into the equation
\eqref{eq:0formsystemc}, or rather into its solution
\begin{align}
R(\tilde f_0^2 - \tilde f_1^2 + 1) = \mbox{const.}
\end{align}
Since $\tilde f_0$ and $\tilde f_1$ are finite, it follows that the
constant vanishes. It turns out that in general this constant is twice the mass
of the black hole, and thus the limit $R\rightarrow 0$ only exists for
the flat geometry. In this case one may easily rewrite
\eqref{eq:0formsystem} and
\eqref{eq:1formsystem}-\eqref{eq:HodgeInterpretation}
such that the variables are $\tilde f_0$, $\tilde f_1$,
$\tilde\alphab := R\alphab$, $\tilde\betab := R\betab$,
$\thetab^0$, $\thetab^1$ and $\omegab$.

We are now in a position to apply the discretisation procedure as
explained in~\cite{bossavit1998:_discr_em_prob}. However, the
situation in GR is significantly different from electrodynamics so
that there is no straightforward implementation. For one thing the
distinction between direct and dual forms is used in electrodynamics
in an elegant way by employing a dual mesh. It is not so clear whether
one can make use of this also in GR, because the description of
dual forms on the dual mesh strongly relies on the Euclidean character
of the metric. Consequently in electrodynamics one only discretises space
using the method of discrete differential forms while we aim at fully
discrete schemes.

Furthermore, electrodynamics is a linear theory while the nonlinearity
of GR is apparent from the appearance of the wedge product which has
to be implemented on the discrete level. So we cannot simply adapt the
implementation of electrodynamics, but we will have to look at the
forms differently. The aim is to split a possibly large system of
nonlinear equations into many small systems of equations in order to
get one independent system for every simplex. How this is done in
practice, is described
in~\cite{frauendiener2006:_discr_differ_forms_gener_relat} and briefly
in the following section.

\section{Implementation of the discrete equations}
\label{discrete_form_section}

In section \ref{spherical_Sparling_equations} and appendix
\ref{app:spher_symm} we derived systems of exterior equations on~$\cM_1$.
Now we present methods for discretising them and
develop numerical schemes. Since there is no unique natural way, we
use different discretisation schemes to explore various possibilities.
The first scheme will be based on the system \eqref{eq:0formsystem},
the second one is obtained through a discretisation of the equations
\eqref{eq:1formsystem},\eqref{eq:algebr} and to get the third scheme we
discretise the system \eqref{eq:1formsystem},\eqref{eq:HodgeInterpretation}.

\subsection{Application of Whitney forms}

As indicated in the introduction, we approximate the manifold by
taking into account finitely many of its subsets.
Moreover we want to use discrete differential forms as an
approximation for the continuous differential forms.

It is known that the opposite, i.e. the extension of a discrete
differential form to a continuous differential form can be done with
the help of the so-called Whitney
forms~\cite{whitney1957:_geomet_integ_theor}. These are a special
class of forms which can be used to construct continuous forms from
discrete forms. However, they can exist only on special domains such
as simplices, $n$-dimensional cubes and shapes, that can be
constructed from a cube by collapsing some of its
edges~\cite{gradinaruhiptmair1999:_whitn} such as pyramids or prisms.
The numerical variables are then the integrals of the forms over the
corresponding figures.

We have chosen the first possibility, i.e. we are searching for an
approximation of space-time by taking subsets into account, that can
be continuously mapped to simplices. In the 2-dimensional case these
simplices are nodes, edges and faces. The reason for this choice
is, that for other shapes it is not possible to get an exterior
product from (anti-)symmetry requirements alone. This leads to
anaesthetic ambiguities, and since the exterior product of Whitney forms
is in general not a Whitney form, symmetry assumptions
are the best way to introduce the discrete exterior product.

Using simplices one gets, up to a normalisation, an exterior product
essentially from the following requirements
\begin{enumerate}
\item The discrete exterior product fulfils the usual commutation rule
  for forms, i.e. for a $p$-form $\alphab^p$ and a $q$-form
  $\betab^q$ we have 
  \begin{equation*}
    \alphab^p\wedge\betab^q = (-1)^{pq}\betab^q\wedge\alphab^p.
  \end{equation*}
\item When the orientation of the simplex is changed, the sign of the
  corresponding value of the discrete exterior product changes. The same is
  true for every Discrete Differential Form.
\end{enumerate}
These requirements lead almost immediately to the following formula
for the discrete exterior product between 1-forms~$\alphab$ and~$\betab$
\begin{equation}
  \label{eq:ext_prod}
\begin{aligned}
  \alphab\betab[n_0,n_1,n_2] &=
  \frac{1}{6}\big(\alphab[n_0,n_1]\betab[n_0,n_2] -
  \alphab[n_0,n_2]\betab[n_0,n_1] \\
   &\qquad+
  \alphab[n_1,n_2]\betab[n_1,n_0] -
  \alphab[n_1,n_0]\betab[n_1,n_2]\\
  &\qquad +
  \alphab[n_2,n_0]\betab[n_2,n_1] -
  \alphab[n_2,n_1]\betab[n_2,n_0]\big),
\end{aligned}
\end{equation}
where the expression $\gammab[n_0,\ldots,n_p]$ is the numerical
variable corresponding to the integral of the $p$-form $\gammab$
over the simplex with nodes $\{n_0,\ldots,n_p\}$ and orientation
given by the ordered tuple of vectors $(n_1-n_0,\ldots,n_p-n_0)$. It
turns out that this definition and its analogues for higher degree
forms yield an algebraic structure which is not associative in
contrast to the continuous case. How this non-associativity influences
the method is not clear. It is clear however, that the terms which
become ambiguous due to the non-associativity are of higher order so
they converge to zero faster in the continuum limit.

For the discretisation of the exterior derivative, we remember Stokes
theorem and get for a 1-form $\alphab$
\begin{equation}
  \label{eq:ext_der}
  \dd\alphab[n_0,n_1,n_2] = \alphab[n_1,n_2]
  -\alphab[n_0,n_2] +\alphab[n_0,n_1]. 
\end{equation}

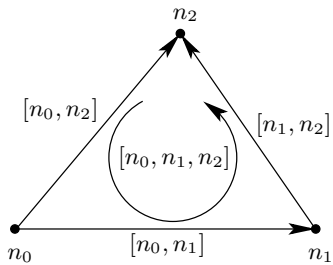
\begin{figure}[htbp]
\centering
\input{Stokes.pstex_t}
\caption{The boundary of the 2-dimensional oriented face $[n_0,n_1,n_2]$ is
composed of three oriented edges $[n_0,n_1]$, $[n_0,n_2]$, $[n_1,n_2]$ and
three nodes $n_0$, $n_1$, $n_2$.}
\end{figure}

\subsection{Properties of the simplicial mesh}
\label{properties}
Now we come to the numerical schemes. Common to all three schemes is
the way of generating a simplicial approximation of a subset of~$\cM_1$.
We start from appropriate initial data. That means from somewhere we
have a 1-dimensional simplicial complex $\mathcal
C_i$~\cite{munkres1993:_simpl_compl_simpl_maps}, that approximates a
space-like curve in~$\cM_1$ (see section~\ref{scenarios} for details).
At each node of $\mathcal C_i$ two linearly independent light-like
directions $\mathbf l_0$ and $\mathbf l_1$ exist. It is clear, that a
light-like curve with tangent-vector $\mathbf l_0$ at a node $n_0$
will have an intersection with a light-like curve with tangent vector
$\mathbf l_1$ at another node $n_1$.\footnote{At least when $n_0$ and
  $n_1$ are sufficiently close to each other.}  To create a face of
the mesh, that contains an edge of $\mathcal C_i$, we require the two
missing edges to be approximations of these light-like curves. Their
intersection becomes the new node $n_2$.  For obvious reasons, these
faces are called \emph{upwards directed}.

This construction seems to be the simplest invariant method to define the
position of $n_2$ in $(1+1)$-dimensional manifolds. It is a geometric 
construction and can be generalised to higher dimensions. The choice of the
nodes at a later time and their connections to the nodes at the initial time
is essentially arbitrary and only restricted by topological considerations.
It is only when the $\thetab^i$ are known on all the edges that the geometry
of the mesh is determined. We will see later that a part of these values
can be specified freely while the rest is determined from the equations.

As in all numerical simulations degeneracies may occur. For instance two
adjacent nodes in the same level may have a time-like distance. However,
this must be seen as a sign that the mesh is too coarse and should be refined.

The other type of faces is called \emph{downwards directed}, and is
created by joining the intersections of the light-like curves in
adjacent upwards directed faces. When $\mathcal C_i$ has $n$ edges, we
now have $n$ upwards directed and $(n-1)$ downwards directed faces.
The collection of these faces will be called the first
\emph{time-step}.

\begin{figure}
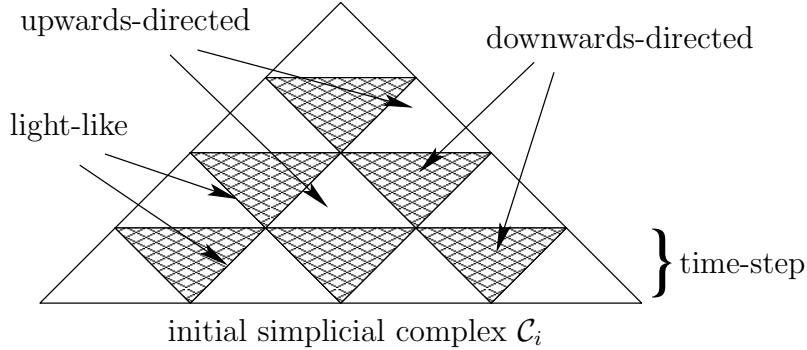

  \begin{center}
    \input mesh.pstex_t
    \caption{The triangulation of the domain of dependence of $\mathcal C_i$.}
    \label{fig:triangulation}
  \end{center}
\end{figure}

Obviously, this procedure can be continued by taking the collection of
the non light-like edges of the downwards directed faces as a new
initial complex $\mathcal C^\prime_i$, until the intersection of the
light-like curves from the boundary of $\mathcal C_i$ is reached. That
means that we calculate the domain of dependence of $\mathcal
C_i$. In principle we could have implemented boundary conditions, but
we wanted to concentrate on the time evolution scheme and periodic boundary
conditions are not possible in spherical symmetry.
Figure~\ref{fig:triangulation} shows the
triangulation.

Having a simplicial mesh, the exterior product and derivative, we can
now take care of the discrete equations.

In what follows we will present three numerical schemes. Common to all
schemes is that for each triangle a system of equation has to be solved.
These are coupled non-linear algebraic
equations. Their analysis is somewhat complicated and
their status is not yet clear. They might not have a unique solution.
However, at least one solution can be found by Newton's iteration
method.  We used the GNU Scientific Library, especially the
implementation of a root finding algorithm called modified Powell
method by the developers \cite{powell1970,gsl2005:_refer_manual}.

\subsection{Scheme I}
\label{subsec:scheme1}

For the first scheme the system (\ref{eq:0formsystem}) is used.
The variables are the discrete
1-forms $\thetab^0$, $\thetab^1$ and $\omegab$ as well as the
discrete 0-forms $f_0$, $f_1$ and $R^{-2}$.  For the upwards directed
faces the numbers
\[
\{f_0[n_0],f_0[n_1],f_1[n_0],f_1[n_1],R^{-2}[n_0],R^{-2}[n_1],
\thetab^0[n_0,n_1],\thetab^1[n_0,n_1],\omegab[n_0,n_1]\}
\] 
are given initial data, and
\begin{align*}
 \{&f_0[n_2],f_1[n_2],R^{-2}[n_2],\\
 &\thetab^0[n_0,n_2],
  \thetab^0[n_1,n_2],\thetab^1[n_0,n_2],\thetab^1[n_1,n_2],\omegab[n_0,n_2],
  \omegab[n_1,n_2]\}
\end{align*}
are the unknowns. 

The equation \eqref{eq:0formsystemc} is the statement that the
function
\begin{align}
  F:=\left(f_0^2-f_1^2+R^{-2}\right)/\left(R^{-2}\right)^{3/2}
\end{align}
is constant on $\mathcal M_1$, and it is implemented in this way.
Thus, we calculate the constant $C:=F[n_0]$ from the (known) values
$f_0[n_0]$, $f_1[n_0]$ and $R^{-2}[n_0]$ and then require
\begin{align}
  \label{eq:constFunc}
  F[n_2]=
  \left(\left(f_0^2-f_1^2+R^{-2}\right)/\left(R^{-2}\right)^{3/2}\right)[n_2]
  =C,
\end{align}
since equation \eqref{eq:0formsystemc} implies that
$F[n_0]=F[n_1]=F[n_2]=C$.

Therefore, the number of equations is six (two 1-form equations for
the two light-like edges, one 2-form equation and
(\ref{eq:constFunc})), but the number of unknowns is nine.  We
eliminate two unknowns by using the definition of the position of
$n_2$ (see section \ref{properties}). We get
\begin{align}
  (\thetab^0-\thetab^1)[n_0,n_2]&=0,&(\thetab^0+\thetab^1)[n_1,n_2]&=0,
\end{align}
expressing the fact that the edges are null.

What remains is the freedom of choosing a gauge. This corresponds to the choice
of a cobasis at $n_2$. The dyad at $n_2$ can be obtained from parallel
transport of $\{\thetab^0,\thetab^1\}$ from the initial
hypersurface to $n_2$ along an edge. Since parallel transport is
defined by $\omegab$ we choose it such that
\begin{align}
\label{eq:gauge_schemeI}
  \omegab[n_0,n_2]&=0.
\end{align}
The reason to choose this gauge condition is of course that it is the most
simple one. Clearly, in general it is the aim to choose the gauge in some
sense `optimal', in order to get good results. However, until now we do not
understand well, what `optimal' means here.
Probably this may become clear once the properties of the equations are better
understood.

In the continuum limit \eqref{eq:gauge_schemeI} corresponds to a dyad that is
obtained through parallely transporting the (known) basis at the initial
hypersurface along the light-like curves with tangent vector
$(\mathbf e_0+\mathbf e_1)$. This fixes the so-called strong conformal
geometry of the null-hypersurface generated by
the spherical family of light rays (see Penrose \cite{PenroseRindler}).

For the downwards directed faces the situation is easier. The 0-form
equation (\ref{eq:constFunc}) is automatically fulfilled at all nodes,
and the 1-form equations \eqref{eq:0formsystema} and \eqref{eq:0formsystemb}
are fulfilled at the
light-like edges. What
remains are two 1-form equations \eqref{eq:0formsystema},
\eqref{eq:0formsystemb} and one 2-form equation \eqref{eq:0formsystemd}.
The unknowns
are the integrals of the 1-forms $\{\thetab^0, \thetab^1,
\omegab\}$ along the new edge.

Altogether we have six equations and six unknowns for the upwards
directed faces, as well as three equations and three unknowns for the
downwards directed ones. That means to obtain a numerical solution we
only have to solve a system of six equations for each upwards directed triangle
and a system of three equations for each downwards directed one.
In the simulations the root finding algorithm sometimes detects a solution that
is a bad approximation of the analytic solution. However, this can be resolved
by choosing other starting values for the Newton iteration, and we did
not investigate it further.

\subsection{Scheme II}
\label{sec:scheme2}
In the second scheme we use the system (\ref{eq:1formsystem}) together with
\begin{align}
  \label{eq:wedgehodge}
  \alphab\tilde{\thetab}^0
  +\betab\tilde{\thetab}^1&=0,&
  \alphab\tilde{\thetab}^1
  +\betab\tilde{\thetab}^0&=0.
\end{align}
In this case, the variables are the discrete 1-forms $\alphab$,
$\betab$, $\tilde{\thetab}^0$, $\tilde{\thetab}^1$ and
$\omegab$.

The given initial data for an upwards directed face are
\begin{align}
\nonumber
  \{\alphab[n_0,n_1],\betab[n_0,n_1],\tilde{\thetab}^0[n_0,n_1],
  \tilde{\thetab}^1[n_0,n_1],\omegab[n_0,n_1]\},
\end{align}
and the unknowns
\begin{align}
  \nonumber
  \{&\alphab[n_0,n_2],\alphab[n_1,n_2],
  \betab[n_0,n_2],\betab[n_1,n_2],
  \tilde{\thetab}^0[n_0,n_2],\tilde{\thetab}^0[n_1,n_2],\\
  \nonumber
  &\tilde{\thetab}^1[n_0,n_2],\tilde{\thetab}^1[n_1,n_2],
  \omegab[n_0,n_2],\omegab[n_1,n_2]\}.
\end{align}
The discretisation of the 2-form equations \eqref{eq:1formsystem},
\eqref{eq:wedgehodge} leads to seven relations.
The number of unknowns is ten. With the same procedure as in scheme~I,
i.e. using the fact that the new edges are light-like and choosing a gauge with
$\omegab[n_0,n_2]=0$, we reduce the number of unknowns to seven.

In this case the downwards directed faces are more difficult to treat.
Since all equations came from the discretisation of 2-forms, we still
have seven equations for these faces, but the number of unknowns is
only five (the values of five 1-forms on the upper edge).  To
get around this difficulty the following idea is used.

In general there is no exact solution of the discrete equations, because there
are more equations than unknowns.%
\footnote{The reason for this problem is of course
that equations \eqref{eq:wedgehodge} describing the Hodge operator can be
localised at the faces of the mesh. This causes difficulties, because it is
not natural for this operator. On a 2-dimensional manifold the continuous
Hodge makes it possible to identify 1-forms with pseudo 1-forms and visa versa.
Thus it would be more natural to localise this operator at 1-dimensional
submanifolds. However, having no dual mesh here, it is not clear how
to achieve this in general.}
However, possibly one can choose the dyad such that finding an exact
solution is possible. To find this optimal gauge one searches for the exact
solution and the dyad simultaneously. Hence we are using the gauge freedom
to change the number of unknowns.

To clarify the details of this procedure we first want to discuss which gauge
choices can be made. As a starting point serves the discussion of the gauge
issues in \cite{frauendiener2006:_discr_differ_forms_gener_relat}.
There it is argued that in the intersection of two separate regions where
the tetrads are chosen independently exist transition maps which mediate
between the different gauges. Obviously these transition maps have the form
of (position dependent) Lorentz transformations.

\begin{figure}[htbp]
\centering
\includegraphics[width=8cm]{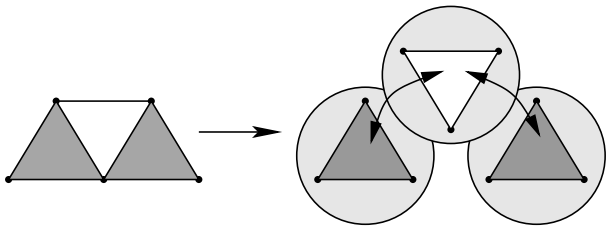}
\caption{Each simplex is viewed as being contained in an open set on which a
tetrad is locally defined. Transition maps mediate between different tetrad
patches and hence between different simplex gauges.}
\label{fig:regauging}
\end{figure}

We may choose an open covering of the manifold with those regions, such that
every open set in the covering contains only a single simplex. Then the
transition maps can be interpreted as gauge transformations from one simplex
to another (cf. figure \ref{fig:regauging}).

In fact we will use these transition maps as the new unknowns, but we will
parameterise them through their action on the tetrad and the connection forms.
Let the dyad and the connection in the upwards directed triangle be
$\{\thetab^0,\thetab^1,\omegab\}$, and in the downwards directed face
$\{\bar{{\thetab}}^0,\bar{{\thetab}}^1,\bar\omegab\}$.
Furthermore let $[n_1,n_2]$ be the common edge of the two simplices.

In two dimensions a Lorentz transformation is completely determined by a
single parameter, its rapidity $\psi$. Hence the action of the transition
function has the form
\begin{align}
\label{eq:transition_map}
\bar\thetab^0&=\cosh\psi\thetab^0+\sinh\psi\thetab^1,&
\bar\thetab^1&=\cosh\psi\thetab^1+\sinh\psi\thetab^0,&
\bar\omegab&=\omegab-\mathbf d\psi.
\end{align}

On the discrete level the rapidity $\psi$ can be seen as a 0-form, i.e.
a map that assigns a number to each node. Yet, in the intersection
of two tetrad patches there are only two nodes $n_1$ and $n_2$, and hence the
discretised transition map is determined by two parameters
$\psi_1=\psi[n_1]$ and $\psi_2=\psi[n_2]$.

Assuming that these parameters are small we may perform a Taylor expansion
of \eqref{eq:transition_map}. The result is that in the leading order
$\thetab^{0/1}[n_1,n_2]-\bar\thetab^{0/1}[n_1,n_2]$
depends on the sum $(\psi_1+\psi_2)$, but
$\omegab[n_1,n_2]-\bar\omegab[n_1,n_2]$ depends on the difference
$(\psi_1-\psi_2)$. Hence we can choose the parameters $\psi_1$ and $\psi_2$
such that the values of $\bar{{\thetab}}^i$ and
${\thetab}^i$ at $[n_1,n_2]$ are the same, while
the values of $\bar{\omegab}$ and ${\omegab}$ differ.

With the two differences of the gauge parameters along the two
light-like edges of the upwards directed face we obtain two new
unknowns. Effectively, this is the same as forgetting about the value
of $\omegab$ at these edges. Hence the two additional unknowns are
the values of $\bar{\omegab}$ at the light-like edges.

It is known from the continuous theory that this regauging should be possible,
but it is not known what kind of restrictions are imposed on the gauge
parameter by the discretisation. Thus, we assume that this procedure
is allowed and check with numerical tests whether this is true.

\subsection{Scheme III}
The third scheme is based on the system
\eqref{eq:1formsystem},\eqref{eq:HodgeInterpretation}.
This scheme is a modification of scheme II in the following sense. In section
\ref{spherical_Sparling_equations} we discussed that
(\ref{eq:wedgehodge}) can be interpreted as $\alphab =
\star\betab$.  When evaluated on the light-like edges of upwards
directed faces, this formula implies
\begin{align}
  \label{eq:discrHodgelight}
  (\alphab-\betab)[n_0,n_2]=0=
  (\alphab+\betab)[n_1,n_2].
\end{align}
These are equations on the two light-like edges.
So, instead of two 2-form equations we have two 1-form equations.

This reinterpretation of (\ref{eq:wedgehodge}) does not change the number
of equations for upwards directed faces, but it does change it 
for downwards directed ones. The new 1-form equations are of course
already satisfied at the light-like edges, so we loose two equations and we
need no regauging, since the number of equations and unknowns are already
equal.

In order to test the influence of the gauge choice at the upwards directed
faces we make another change. We will not use $\omegab[n_0,n_2]=0$, but
\begin{equation}
\label{eq:gauge_implicit_III}
  \omegab[n_0,n_2]+\omegab[n_1,n_2]=0
\end{equation}
instead. This is an implicit and non-local definition of the gauge, but it
has the advantage that the edges $[n_0,n_2]$ and
$[n_1,n_2]$ are treated on a par.

An obvious problem of this scheme is that
the discretisation \eqref{eq:discrHodgelight} of the Hodge-$\star$ operator
can only be applied when two edges of every face are light-like. Clearly this
limits its applicability and the question occurs whether
a similar discrete Hodge operator exists for space-like and time-like
edges. Unfortunately we did not find such a discretisation, but we want
to point out what we believe are the first steps to find a more general
discrete Hodge operator with similar properties as
\eqref{eq:discrHodgelight}.

For the nodes $x$, $y$ and $z$ we may interpret the edges $e=[x,y]$ and
$\bar e = [x,z]$, as vectors in the tangent space at $x$.
The 2-dimensional Hodge operator is defined through the equation
\begin{align}
\label{eq:def_Hodge}
\betab\wedge\star\alphab = g^{-1}(\betab,\alphab)\thetab^0\thetab^1\quad
\forall\betab.
\end{align}
Applying \eqref{eq:def_Hodge} to the ``vectors'' $e$ and $\bar e$ and
identifying the result of applying $\alphab$ and $\star\alphab$ to
these vectors with $\alphab[e]$, $\alphab[\bar e]$, $\star\alphab[e]$ and
$\star\alphab[\bar e]$ appropriately leads to
\begin{align}
  \label{genHodge}
  \left(\begin{array}{c}\star\alphab[e]\\ \star\alphab[\bar e] \end{array}
  \right) &=
  \frac{1}{e^0\bar e^1 -e^1\bar e^0}\left(
    \begin{array}{cc}
      e^1\bar e^1 -e^0\bar e^0 &  e^0 e^0 -e^1 e^1 \\
      \bar e^1\bar e^1 - \bar e^0\bar e^0 & e^0\bar e^0 -e^1\bar e^1 
    \end{array}
  \right)
  \left(\begin{array}{c}\alphab[e]\\ \alphab[\bar e] \end{array}\right),
\end{align}
with $e^i=\thetab^i[e]$, $\bar e^i=\thetab^i[\bar e]$.
This is a possible definition of a discrete Hodge operator that can also
be applied to non light-like edges, and when $e$ or $\bar e$ is
light-like it becomes the discrete Hodge operator (\ref{eq:discrHodgelight}).

However, we identified the edges with vectors at $x$. In principle there is
no reason that e.g. $e$ is a vector at $x$ and not at $y$. But imposing
the analogue of \eqref{genHodge} at $y$ and $z$ leads to an inconsistent
system. Thus the question is how to formulate a discrete geometry in
a consistent way. This will be the topic of future investigations.

\section{Test scenarios}
\label{scenarios}

Having clarified how we apply discrete differential forms in General
Relativity, we now describe how we tested the obtained code. Since the
spherically symmetric vacuum solutions of the Einstein field equations
are all contained within the Kruskal solution \cite{Weinberg} for some value
of the mass parameter $M$ we know the exact solutions for our problem.
The Kruskal metric has in the standard coordinate system
the form
\begin{align}
g=f(R) \left(\mathbf dT\otimes\mathbf dT-\mathbf dX\otimes\mathbf dX\right)
-R^2\left(\mathbf d\vartheta\otimes\mathbf d\vartheta + \sin^2\vartheta\,
\mathbf d\varphi\otimes\mathbf d\varphi\right).
\end{align}
There the functions $f(R)$ and $R=R(T,X)$ are defined through
\begin{align}
\label{eq:def_f}
f(R)&=\frac{32M^3}{R}e^{-\frac{R}{2M}},&
R(T,X) &= 2 M\left(1+ W\left((X^2-T^2)/e\right)\right),
\end{align}
where $W$ is the Lambert
W-function~\cite{corlessgonnet1996:_lamber_w}. To
start the time evolution we need initial data which we obtain from one
of the analytic solutions. These data must be given as edge- and node
values.

\subsection{The continuous forms}

The best way to do this is to find a description of the geometry, that
uses (continuous) differential forms, i.e. maps from the set of
\emph{all} submanifolds to the real numbers. However, such an
(abstract) map can hardly be useful for concrete calculations, because
without coordinates it is even difficult to describe the position of a
point. Therefore we take coordinate representations of the Minkowski and
Schwarz\-schild geometries. For the Schwarz\-schild geometry it is
convenient to use Kruskal coordinates, because then the light-like
curves (which are special for the described method) take a very simple
form and hence it is easier to compare the results.

To obtain the differential forms, we make a gauge choice, i.e. we
choose $\thetab^0$ and $\thetab^1$, such that they generate the
corresponding metric.  In Minkowski space the natural choice is
\begin{equation}
\label{eq:Minkowski_forms}
\begin{aligned}
  \thetab^0&=\dd t,&
  \thetab^1&=\dd r,&
  \alphab&=\frac{\dd r}{r},&
  \betab&=\frac{\dd t}{r},\\
\omegab&=0,&
  R^2&=r^2,&
  f_0&=0,& f_1&=\frac{1}{r},\\
  \end{aligned}
\end{equation}
where $r$ and $t$ are the standard space and time coordinate, respectively.

In Kruskal coordinates we use, with the standard space and time coordinates $X$ and
$T$, as well as the mass parameter $M$,
\begin{subequations}
\label{eq:Kruskal_forms}
\begin{align}
  \thetab^0 &=\sqrt{f(R)}dT,&
  \thetab^1 &=\sqrt{f(R)}dX,\\
  f_0 &=-\frac{T h(R)}{\sqrt{f(R)}},&
  f_1 &= \frac{X h(R)}{\sqrt{f(R)}},\\
  \alphab &=h(R)(XdX-TdT),& \betab
  &=h(R)(XdT-TdX),\\
\label{eq:def_Kruksal_omega_R}
  \omegab &=g(R)(TdX-XdT),&
  R^2 &= 4 M^2 \left(1+ W\left((X^2-T^2)/e\right)\right)^2,
\end{align}
\end{subequations}
where the functions $h$ and $g$ are defined through
\begin{subequations}
\begin{align}
h(R)&=\frac{8M^2}{R^2}e^{-\frac{R}{2M}},\\
g(R)&=\frac{4M^2}{R}\left(\frac{1}{R}+\frac{1}{2M}\right)e^{-\frac{R}{2M}}.
\end{align}
\end{subequations}

\subsection{Getting initial values}

Next an initial hypersurface has to be chosen. For the test we used
curves, whose space-time coordinates $(y^0,y^1)$ depend linearly on the
curve parameter. These curves will be called ``straight'':
\begin{equation}
  \label{eq:def_straight_Kruskal}
  \left(
    \begin{array}{c}
      y^0 \\ y^1
    \end{array}
    \right) =
  \left(
    \begin{array}{c}
      y_0^0 \\ y_0^1
    \end{array}
    \right)
    + \lambda   \left(
    \begin{array}{c}
      y_1^0 \\ y_1^1
    \end{array}
    \right),
\end{equation}
for some $\lambda\in[0,1]$ and fixed $y_0^0,y_0^1,y_1^0,y_1^1$.

We get the initial edges and nodes by subdividing this curve into
pieces of equal `coordinate length'. That is we start at a
hypersurface of the form (\ref{eq:def_straight_Kruskal}) with boundary
at $\lambda = 0$ and $\lambda = 1$ and subdivide it into $n$ pieces.
The pieces are again straight, and $\lambda$ takes values in the
intervals $[(i-1)/n,i/n],\;i=1,\ldots,n$.  On the edges we can
integrate the continuous 1-forms known from the analytic solution, and
the results are the initial values for the corresponding discrete
1-forms.

To get initial values for the 0-forms, we evaluate the corresponding
continuous functions at the boundary points $\lambda =
i/n,\;i=0,\ldots,n$ of the sub-intervals. These values are
invariant under coordinate transformations so that we do not start with
coordinate dependent values in the beginning.

\subsection{Examination of the results}

In order to compare the numerical results with the analytical solution
we need to determine the location in space-time of the nodes and edges
used in the algorithm. This can be a difficult task because in
principle one needs to solve the geodesic equations to obtain the
light rays used to define the nodes in the next time-slice. However,
here this is very much simplified since in Kruskal coordinates as well
as in Minkowski coordinates the radial light rays move on straight
lines.  

When comparing the results we need to worry about the gauge.  I.e.
when we use different gauges for the discrete approximation and for
the analytic solution, we cannot expect to get the same results.
However, if the method is feasible, we can expect that gauge invariant
discrete variables are good approximations of the continuous ones.
Gauge independent values are for instance the lengths
$l=\sqrt{|(\thetab^0)^2-(\thetab^1)^2|}$ of the space-like edges,
the values of the 1-form $\alphab$
on these edges and the values of the function $R^{-2}$ at the nodes.

Another way to evaluate a numerical method is a self-convergence test.
There one compares solutions obtained on coarse meshes with a solution that is
calculated on a fine mesh. It is clear that a method that converges to the
analytical solution when the mesh is refined is self-convergent, too.

\section{Results}
\label{results}

For the number of initial edges we have always chosen a power of two
($n=2^i,\;i=1,2,3,\ldots$), and calculated half of the maximal number
of time-steps: for $n$ initial edges, we calculate $n/2$ time-steps.
The last time-step then contains $n/2$ downwards directed simplices.
Each of these simplices contains one non light-like edge and each of
these edges contains two nodes. Altogether these are $n/2$ edges and
$(n/2+1)$ nodes, since nodes of adjacent edges coincide.

It turns out that the computational costs of the three schemes are quite
comparable. In scheme~I the solution in a mesh with approximately
200.000 faces is obtained in one minute on a 700MHz PC. In schemes~II and III
it takes about 40\% and 10\% longer respectively.
This could be expected, because in scheme I one has to solve systems
of six and three equations for the upwards and downwards directed triangles
respectively. In schemes~II and III the sizes of these systems are seven
and seven, respectively seven and five.
Thus, in scheme~I the number of equations is much smaller. Clearly, since
there is a system of equations on each individual face, the time that
is needed to find the solution depends linearly on the number of faces,
and hence quadratically on the number of initial edges.

In the left diagrams of Figs. \ref{fig:Mink1},
\ref{fig:Kruskal_spacelike} and \ref{fig:Kruskal_timelike} we show the
maxima over the $n/2$ edges of the relative errors in the values of
the 1-form $\alphab$ for schemes II and III as well as the maxima
over the $(n/2+1)$ nodes of the relative errors in the values of
$R^{-2}$ for scheme I.%
\footnote{Since we compare relative errors it is feasible to draw different
quantities in the same diagram. Furthermore $\alphab$ and $R^{-2}$ both
are used as parameterisations of the eliminated degrees of freedoms in the
orbits of the $SO(3)$-group action.}
In the right diagrams of Figs.
\ref{fig:Mink1}, \ref{fig:Kruskal_spacelike} and
\ref{fig:Kruskal_timelike} the maxima over the $n/2$ edges of the
relative error of the invariant lengths $l$ are plotted.

\subsection{Minkowski space-time}

\begin{figure}[htb]
\begin{tabular}{cc}
\includegraphics[scale=.84]{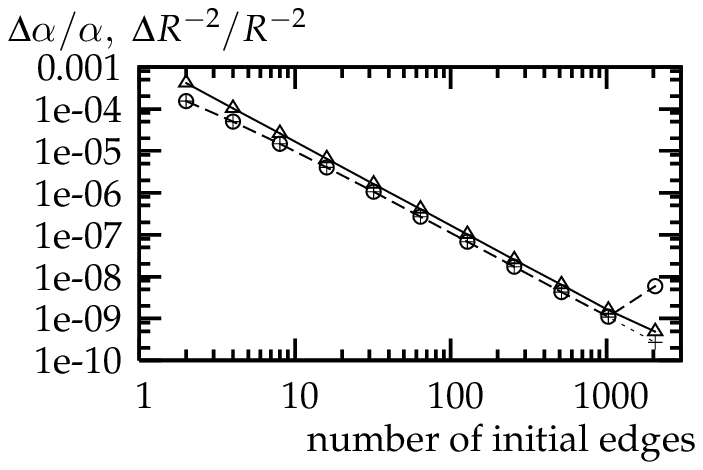}&
\includegraphics[scale=.84]{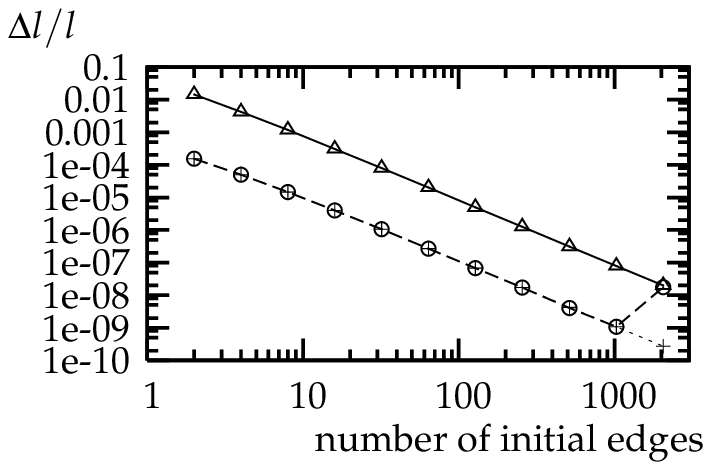}
\end{tabular} 
  \begin{pspicture}(10.5,0.2)
    \psset{linewidth=.6pt}
    \psset{dash=4pt 2pt}
    \psdots[dotstyle=triangle](.5,0.1)
    \psline(0,0.1)(1,0.1)
    \put(1.2,0){scheme~I}
    \psdots[dotstyle=o](4.5,0.1)
    \psline[linestyle=dashed](4,0.1)(5,0.1)
    \put(5.2,0){scheme~II}
    \psdots[dotstyle=+](8.5,0.1)
    \psline[linestyle=dashed,dash=2pt 5pt](8,0.1)(9,0.1)
    \put(9.2,0){scheme~III}
  \end{pspicture}
\caption{Maximal relative errors in Minkowski space-time.
  Left: values of $R^{-2}$ at the nodes (scheme I) and values of
  $\alphab$ at the space-like edges (schemes II,III).
  Right: invariant lengths of space-like edges.}
\label{fig:Mink1}
\end{figure}

For the initial hypersurface we have chosen the set $\{ (t,r) =
(0,1+\lambda) : \lambda\in[0,1]\}$, with the standard time and space
coordinates $t$ and $r$, respectively. We compare the results with the
analytic solution at the hypersurface $\{(t,r) = (0.25,
1.25+\lambda/2) : \lambda\in[0,1]\}$.

We see from figure \ref{fig:Mink1}, that the relative error converges for all
three schemes quadratically to zero when the typical size of simplices in
the mesh is decreased. The error of the lengths is about 100 times bigger for
scheme I, but it remains small even for the coarsest mesh.

\subsection{Kruskal geometry}

In Kruskal geometry we test the code for one space-like and one
time-like initial 
hypersurface. Since in Kruskal coordinates the horizon is a regular
null-hypersurface we can test how the code behaves near the 
event horizon. So, we take the space-like curve to cross the horizon.

\subsubsection{Space-like initial data}
In $(T,X)$-coordinates we choose the initial hypersurface
$\{T=\lambda/2,X=-1+2\lambda,\lambda\in[0, 1]\}$, and compare the results at
the hypersurface $\{T=5/8+\lambda/4,X=-3/8+\lambda,\lambda\in[0, 1]\}$.
\begin{figure}[htb]
\begin{tabular}{ll}
\includegraphics[scale=.84]{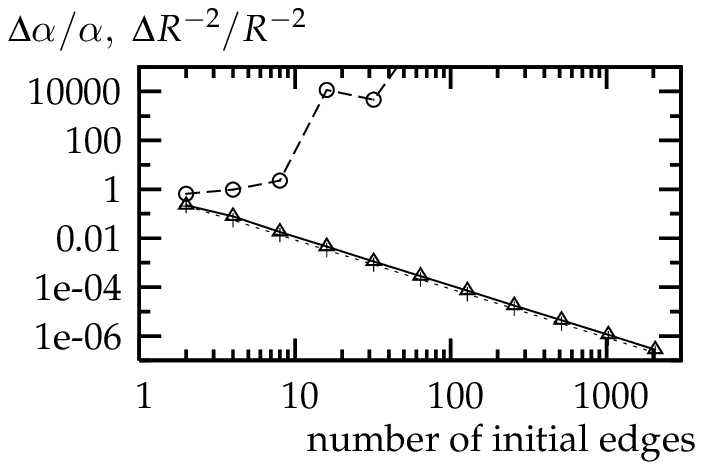}&
\includegraphics[scale=.84]{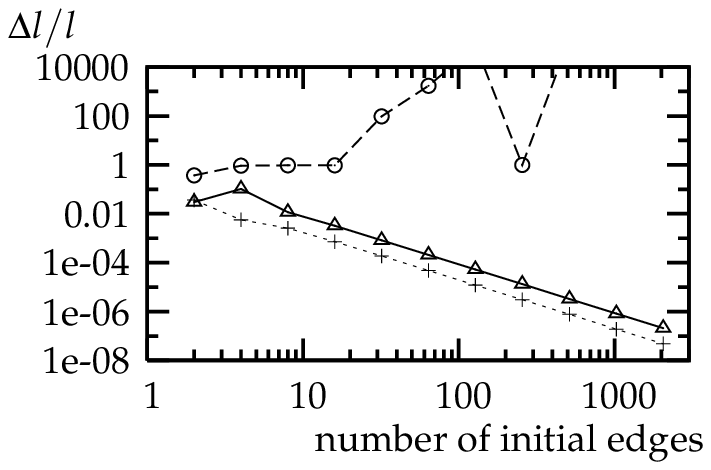}
\end{tabular} 
  \begin{pspicture}(10.5,0.2)
    \psset{linewidth=.6pt}
    \psset{dash=4pt 2pt}
    \psdots[dotstyle=triangle](.5,0.1)
    \psline(0,0.1)(1,0.1)
    \put(1.2,0){scheme~I}
    \psdots[dotstyle=o](4.5,0.1)
    \psline[linestyle=dashed](4,0.1)(5,0.1)
    \put(5.2,0){scheme~II}
    \psdots[dotstyle=+](8.5,0.1)
    \psline[linestyle=dashed,dash=2pt 5pt](8,0.1)(9,0.1)
    \put(9.2,0){scheme~III}
  \end{pspicture}
\caption{Maximal relative errors in Kruskal space-time (space-like initial
data). Left: values of $R^{-2}$ at the nodes (scheme I) and values of $\alphab$
at the space-like edges (schemes II,III). Right: invariant lengths of
space-like edges.}
\label{fig:Kruskal_spacelike}
\end{figure} 

Fig.~\ref{fig:Kruskal_spacelike} shows that the schemes~I and III
provide small errors and the relative error converges quadratically to
zero when the size of the simplices is decreased.  The errors in
scheme II on the other hand are very big, and even become bigger when
the mesh is refined. Clearly this points to a problem in scheme~II that
we discuss later.

\subsubsection{Time-like initial data}
In $(T,X)$-coordinates we choose the initial hypersurface
$\{T=\lambda,X=3,\lambda\in[0, 1]\}$. The comparison of the results is
done at the hypersurface $\{T=0.25+\lambda/2,X=3.25,\lambda\in[0,
1]\}$.
\begin{figure}[htb]
\begin{tabular}{cc}
\includegraphics[scale=.84]{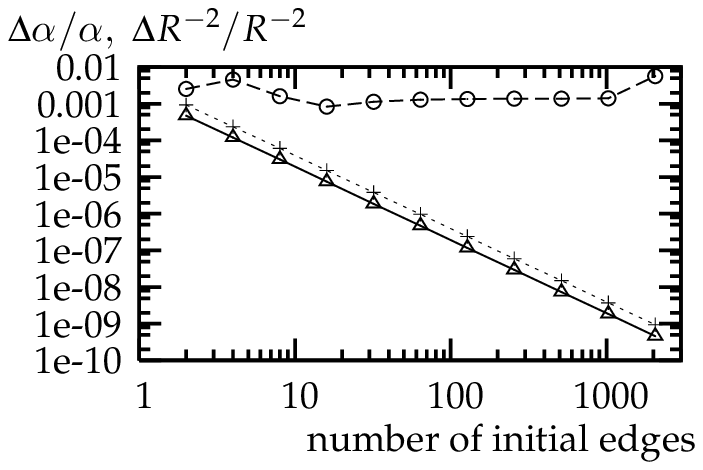}&
\includegraphics[scale=.84]{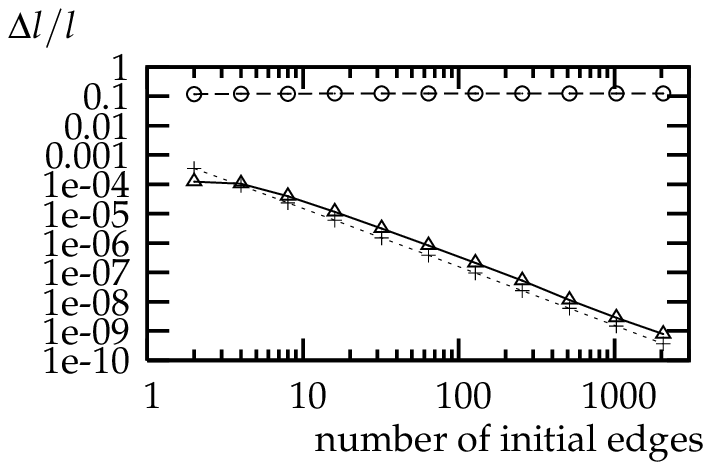}
\end{tabular} 
  \begin{pspicture}(10.5,0.2)
    \psset{linewidth=.6pt}
    \psset{dash=4pt 2pt}
    \psdots[dotstyle=triangle](.5,0.1)
    \psline(0,0.1)(1,0.1)
    \put(1.2,0){scheme~I}
    \psdots[dotstyle=o](4.5,0.1)
    \psline[linestyle=dashed](4,0.1)(5,0.1)
    \put(5.2,0){scheme~II}
    \psdots[dotstyle=+](8.5,0.1)
    \psline[linestyle=dashed,dash=2pt 5pt](8,0.1)(9,0.1)
    \put(9.2,0){scheme~III}
  \end{pspicture}
\caption{Maximal relative errors in Kruskal space-time (time-like initial
data). Left: values of $R^{-2}$ at the nodes (scheme I) and values of $\alphab$
at the space-like edges (schemes II,III). Right: invariant lengths of
space-like edges.}
\label{fig:Kruskal_timelike}
\end{figure} 

Again, as can be seen in Fig.~\ref{fig:Kruskal_timelike},
schemes~I and~III provide very small errors that converge quadratically to
zero when the simplex size is reduced. With 1\% for $\alphab$ and
10\% for the lengths the errors of scheme~II are in this case also quite small.
However, the relative errors do not become smaller for finer meshes.

Since in fig. \ref{fig:Kruskal_timelike} one sees that the errors of the
results in fine meshes are nearly the same as the errors in coarse
meshes, we also performed a self-convergence test, where the number of
initial edges in the finest mesh was 2048. The result is that
scheme~II converges linearly to a solution that differs from the
analytical one.

\section{Discussion}
\label{discussion}

From the results presented in the last section we conclude that the
schemes~I and III provide very good results. Especially scheme~III converges
quadratically to the analytical solution in every simulation we performed.
Also in scheme~I most of the simulations led to quadratically convergent
results. There are only a few regions in Kruskal space-time where the
errors of this scheme become quite large. This is the the case for
$R\approx 3M$.  The reason for this is, that there two
solutions of the discrete equations are close to each other. The root
finding algorithm then sometimes chooses the wrong one.  However, this
should be solvable with an optimised choice of starting values for the
Newton iteration, but we did not find a practical way to obtain such
starting values.

The errors obtained with scheme~II are on the other hand
quite large. This scheme is in some situations not convergent and its
behaviour strongly depends on the initial data.
We conclude that it is not feasible for numerical
calculations and thus seems to be ill-posed. Now, the question is what
the reasons for these problems are.

The difference between the schemes II and III was the direct
implementation of the Hodge operator in scheme~III while scheme~II
made use of the equations~(\ref{eq:wedgehodge}), which have been added
to the system in order to encode the duality between the forms
$\alphab$ and $\betab$. It turns out, that far from the horizon
in the exterior the values of $\alphab$ on the space-like edges
are much smaller than the values of $\betab$, while far from the
horizon in the interior it is the other way round. Near the horizon the
two 1-forms interchange their role in this sense,
and are hence of comparable size.

Near the horizon the solution of (\ref{eq:wedgehodge})
largely differs from $\alphab=\pm\betab$. This property appeared in all
simulation where the initial hypersurface crossed the horizon, especially
it seems to be independent of the location of the nodes in the initial
hypersurface. The origin of this behaviour seems to be the ill-conditioning
of the linear system corresponding to (\ref{eq:wedgehodge}). Thus
(\ref{eq:wedgehodge}) is a bad implementation of the
Hodge operator in that region.

Clearly the Hodge operator \eqref{eq:discrHodgelight} leads to a convergent
scheme. However, \eqref{eq:discrHodgelight} was inspired by the dual mesh of
electrodynamics. In a $(1+1)$-dimensional Lorentz geometry it is
convenient to define the dual of a light-like edge to be the edge itself.

Yet, we wanted to avoid the construction of a dual mesh. The first
reason for this decision is that without a dual mesh the discrete
system decouples into small systems for every face.  The second reason
is that the dual of an initial edge does not lie in the initial
hypersurface and hence we do not know how to specify initial values
for it.  The third reason is that the dual cells are in general not
simplices anymore, which causes problems in defining a natural discrete
exterior product without ambiguities.

The discrete Hodge operator (\ref{eq:wedgehodge}) also made it necessary to
use the regauging described in section \ref{sec:scheme2}.
Although we cannot make definite statments about this procedure yet, it seems
that it is also a source of errors in scheme~II. These problems
probably arise because the notion of a gauge transformation for the
discrete equations has not been clarified completely so far.
This is planned to be the topic of future investiagtions.

\section{Conclusion}
\label{Conclusion}

In this article we presented first results of the application of
discrete differential forms in General Relativity. It was shown, that
the method is quite promising. Several schemes were found whose
results are close to the analytic solution and the errors of which
converge quadratically to zero.

We discussed that one has to be careful with the definition of
discrete Hodge operators and that the notion of discrete gauge
transformations is not completely understood.
Making a wrong decision in these fields
can lead to results with big errors.

Now one has to get a better understanding of gauge transformations for
the discrete equations,
a more general definition of the Hodge
operator is necessary to be able to include matter fields and the
method must be applied in space-times with smaller symmetry groups, in
order to get physically more relevant solutions.

\section*{Acknowledgments}
We want to thank A. Bossavit for very helpful discussions and
suggestions about the structure of discrete manifolds.  This work is
supported by the Deutsche Forschungsgemeinschaft within the SFB 382 on
``Methods and algorithms for the simulation of physical processes on
high performance computers''.

\appendix

\def\HH{\mathbb{H}}
\def\VV{\mathbb{V}}

\section{Derivation of the reduced system}
\label{app:spher_symm}

\subsection{Spherical symmetry}
\label{sec:spherical-symmetry}

In this appendix we will derive the reduced exterior
system~\eqref{eq:0formsystem}. So we assume the existence of an
isometric action of the rotation group~$G=SO(3)$ on the space-time
manifold i.e., for each element $a\in G$ there is an isometry $\phi_a$
of $(\cM,g)$. We assume that the orbits of this action are 2-dimensional
submanifolds except for the fixed points, which form a 1-dimensional
submanifold, called the origin $\mathcal O$.
Clearly, the 2-dimensional orbits are
round spheres, they carry an induced metric which is a constant
multiple of the unit-sphere metric.

Given a point $p\in \cM\setminus\mathcal O$ and the orbit $S_p$ through $p$
we can split $T_p\mathcal M$ into the 2-dimensional
tangent space $\HH_p$ of $S_p$ at $p$, which we call the \emph{horizontal}
space and its orthogonal complement $\VV_p$, the \emph{vertical} space.%
\footnote{For a point $p\in\mathcal O$ this decomposition of $T_p\mathcal M$
is not possible, but this is no problem, because in the simulations
the origin was not included in the computational domain.}
The isometric action maps vertical spaces into vertical ones and
horizontal spaces into horizontal ones. Fix a point $p$ then the isotropy
group of $p$ is isomorphic to $SO(2)$ and the symmetry action defines
a representation of $SO(2)$ on $\HH_p$. This is isomorphic to the
defining representation. On the other hand,
the isotropy group of $p$ acts trivially on $\VV_p$. This follows from
the fact that such an action defines a homomorphism from $SO(2)$ to
$SO(1,1)$. Due to the different topologies of these groups this must
be trivial.

We will be concerned with invariant objects, i.e., objects which are
mapped onto themselves by this symmetry action. Consider an invariant
function $f$. It satisfies $\phi_a^*f=f\circ \phi_a=f$ for every
rotation $a\in G$. This implies that $f$ must be constant on each
orbit because for any two points on a given orbit there is a rotation
which maps one to the other. 

An invariant vectorfield $\mathbf{V}$ coincides with each
push-forward, i.e., $\phi_{a*}\mathbf{V}=\mathbf{V}$. Suppose
$\mathbf{V}$ is horizontal then it follows from the free action of the
isotropy groups on the horizontal spaces that $\mathbf{V}$ must indeed
vanish because at each point $p$ it must coincide with all its images
under elements of the isotropy group. Thus, an invariant vectorfield
cannot have horizontal components so it is always vertical.  It
follows from this that the covariant derivative of an invariant
vectorfield along an invariant vectorfield is again invariant, hence
vertical. Also the commutator of two invariant vectorfields is
vertical. This implies that the subbundle of the tangent bundle
consisting of vertical vectors forms an integrable distribution: any
vertical vectorfield $\mathbf{U}$ can be written as a linear
combination of invariant vectorfields $\mathbf{X}_1$ and $\mathbf{X}_2$
\begin{align}
\mathbf{U} = U^1 \mathbf{X}_1 + U^2 \mathbf{X}_2
\end{align}
so that the commutator of two such vectorfields is
\begin{align}
\left[ \mathbf{U}, \mathbf{V}\right] = \mathbf{U}(V^i) \mathbf{X}_i -
\mathbf{V}(U^i) \mathbf{X}_i + U^iV^k\left[\mathbf{X}_i, \mathbf{X}_k\right],
\end{align}
hence it is vertical. The maximal integral manifolds will be called
vertical surfaces. This discussion shows that the space-time has the
topology $\cM=\cM_1 \times S^2$ where $\cM_1$ is a two-dimensional
manifold. We define the two canonical projections
\begin{align}
\pi: \cM \to \cM_1,\qquad \rho: \cM \to S^2
\end{align}
mapping on the first, resp. second factor.

We can set up adapted local coordinates as follows. Fix a point $p$
and assign to it the coordinates $(x^0,x^1,x^2,x^3)$ where $(x^0,x^1)$
are local coordinates near $\pi(p)\in \cM_1$ and
$(x^2,x^3)$ are local coordinates near $\rho(p)\in S^2$. In these
coordinates the projections are $\pi(x^0,x^1,x^2,x^3)=(x^0,x^1)$ and
$\rho(x^0,x^1,x^2,x^3)=(x^2,x^3)$. 

We recall the following facts. The projections $\pi$ and $\rho$ induce
isomorphisms between $\VV_p$ and $T_{\pi(p)}\cM_1$ resp. $\HH_p$ and
$T_{\rho(p)}S^2$. The group action happens only on the second factor,
i.e., $\pi \circ \phi_a = \pi$. This implies that $p$-forms on $\cM$
which are pull-backs from $\cM_1$ are invariant and annihilate
horizontal vectors. Conversely, a
$p$-form on $\cM$ which is invariant and annihilates horizontal
vectors is the pull-back of a form on $\cM_1$. This is true for any
multilinear map.
Furthermore, invariant vectorfields (which are necessarily vertical)
project down to a well-defined vectorfield on $\cM_1$ and each
vectorfield on $\cM_1$ corresponds to an invariant
vectorfield on $\cM$.

The metric on $\cM$ can be written in the form
\begin{align}
g = g_1 + g_2
\end{align}
where $g_1$ ($g_2$) is non-zero only for vertical (horizontal)
vectors. This shows that the topological decomposition is also an
orthogonal decomposition. Furthermore, for any infinitesimal isometry
$\xi$ we have
\begin{align}
0 = \cL_\xi g = \cL_\xi g_1 + \cL_\xi g_2
\end{align}
and the orthogonality properties imply that the two terms on the right
have to vanish separately,
\begin{align}
\cL_\xi g_1 = 0 = \cL_\xi g_2.
\end{align}
Now, the facts that $\cL_\xi g_1=0$ and that $g_1(X,\cdot)$ vanishes
for any horizontal vector $X$ imply that $g_1$ is the pull-back of a
metric $h$ on $\cM_1$. 

The metric $g_2$ is conformal to the metric $\delta$ on the unit
sphere $S^2$ where the conformal factor may depend on the coordinates
$(x^0,x^1)$. Thus, we can write the space-time metric in the form of a
warped product
\begin{align}
g = \pi^* h + R^2 \rho^* \delta \qquad \text{where } R^2: \cM_1 \to \RR
\end{align}

We can now set up an adapted tetrad $(\thetab^i)_{i=0:3}$. To this end
we choose a frame 
$(\varthetab^0,\varthetab^1)$ on $(\cM_1,h)$ and a frame
$(\varthetab^2,\varthetab^3)$ on $(S^2,\delta)$ and set
\begin{align}
(\thetab^0,\thetab^1,\thetab^2,\thetab^3) =
(\pi^*\varthetab^0, \pi^*\varthetab^1, R\,\rho^*\varthetab^2, R\,\rho^*
\varthetab^3).
\end{align}
Using this tetrad in the first structure equation one finds after some
calculation that the connection forms are
\begin{equation}
\begin{aligned}
  \omegab^0{}_1 &= \pi^* \omegab,\quad& 
  \omegab^2{}_0 &= f_0 \thetab^2,\quad& 
  \omegab^3{}_0 &= f_0 \thetab^3,\\
  \omegab^2{}_3 &= \rho^* \varpib,   & 
  \omegab^2{}_1&=f_1 \thetab^2,&
  \omegab^3{}_1 &= f_1 \thetab^3.
\end{aligned}\label{eq:conn_forms}
\end{equation}
Here, $\omegab$ resp. $\varpib$ are the connection forms of the metrics
on $\cM_1$ resp. $S^2$ and 
\begin{align}
\label{eq:def_alpha}
\alphab\equiv f_0\thetab^0 + f_1\thetab^1 = \dd R/R
\end{align}
is an invariant 1-form on $\cM_1$.

\section{Reduction to 1+1 dimensions}
\label{app:reduction}

Our goal is to express the field equations~(\ref{eq:vacuum_equations}) as
equations on $\cM_1$. The easiest way to achieve this goal is to
compute the Nester-Witten and Sparling forms from the connection
forms~\eqref{eq:conn_forms}. This yields the following result
\begin{subequations}
  \label{eq:L_simple}
  \begin{align}
  \label{eq:L02_simple}
    L_0 &= 2f_1 \thetab^2\thetab^3 - \omegab^2{}_3 \thetab^1,&
    L_2 &= -f_1 \thetab^0\thetab^3 - f_0 \thetab^1\thetab^3 -
    \omegab^0{}_1 \thetab^3,\\
  \label{eq:L13_simple}
    L_1 &= 2f_0 \thetab^2\thetab^3 + \omegab^2{}_3 \thetab^0,&
    L_3 &= f_1 \thetab^0\thetab^2 + f_0 \thetab^1\thetab^2 +
    \omegab^0{}_1 \thetab^2
  \end{align}
\end{subequations}
for the Nester-Witten form, while the Sparling 3-form is given by
\begin{subequations}
\label{eq:S_simple}
\begin{align}
\label{eq:S0_simple}
    S_0 &= 2 f_0 f_1 \thetab^0\thetab^2\thetab^3 + (f_0^2 + f_1^2)
    \thetab^1\thetab^2\thetab^3 + 2f_0 \omegab^0{}_1\thetab^2\thetab^3
    - \omegab^2{}_3 \omegab^0{}_1 \thetab^0,\\
\label{eq:S1_simple}
    S_1 &= 2 f_0 f_1 \thetab^1\thetab^2\thetab^3 + (f_0^2 + f_1^2)
    \thetab^0\thetab^2\thetab^3 + 2f_1 \omegab^0{}_1\thetab^2\thetab^3
    + \omegab^2{}_3 \omegab^0{}_1 \thetab^1,    \\
\label{eq:S2_simple}
    S_2 &= f_0 \omegab^0{}_1 \thetab^0 \thetab^3 - f_0 \omegab^2{}_3
  \thetab^1 \thetab^2 + f_1 \omegab^0{}_1 \thetab^1\thetab^3 - f_1
  \omegab^2{}_3 \thetab^0 \thetab^2 + \omegab^0{}_1 \omegab^2{}_3 \thetab^2,\\
\label{eq:S3_simple}
  S_3 &= -f_0 \omegab^0{}_1 \thetab^0 \thetab^2 - f_0 \omegab^2{}_3
  \thetab^1 \thetab^3 - f_1 \omegab^0{}_1 \thetab^1\thetab^2 - f_1
  \omegab^2{}_3 \thetab^0 \thetab^3 + \omegab^0{}_1 \omegab^2{}_3 \thetab^3.    
\end{align}
\end{subequations}
Let us now compute the exterior derivative of $L_0$. We obtain
\begin{align}
\dd L_0 = 2\dd f_1 \thetab^2\thetab^3 + 2 f_1 \dd(\thetab^2\thetab^3)
- \dd \omegab^2{}_3\thetab^1 + \omegab^2{}_3 \dd\thetab^1.
\end{align}
Since $\thetab^2\thetab^3=R^2\rho^*(\varthetab^2\varthetab^3)$ we find 
\begin{align}
\dd(\thetab^2\thetab^3) = \frac{\dd R^2}{R^2} \thetab^2\thetab^3
\end{align}
and for $\dd \omegab^2{}_3$ we obtain
\begin{align}
\dd \omegab^2{}_3 = \rho^*\dd \varpib = \rho^*\Omegab
\end{align}
where $\Omegab$ is the curvature 2-form of the unit sphere which is
given by $\Omegab=\varthetab^2\varthetab^3$. Hence,
\begin{align}
\dd\omegab^2{}_3 = \rho^*(\varthetab^2\varthetab^3) = \frac1{R^2}
\thetab^2\thetab^3.
\end{align}
Furthermore, making use of the structure of the connection forms we
have 
\begin{align}
\dd\thetab^1 = -\omegab^1{}_0\thetab^0 -\omegab^1{}_2\thetab^2
-\omegab^1{}_3\thetab^3 = -\omegab^1{}_0\thetab^0  .
\end{align}
This shows that with $\thetab^1$ also $\dd\thetab^1$ is invariant
(this also follows from the fact that $\thetab^1=\pi^*\varthetab^1$ so
that $\dd\thetab^1=\pi^*\dd\varthetab^1$).
Taken together, we have
\begin{equation}
\label{eq:dL0_appB}
\begin{aligned}
  \dd L_0 &= \left(2\dd f_1 + 2 f_1 \frac{\dd R^2}{R^2} - \frac1{R^2}
  \thetab^1 \right)\thetab^2\thetab^3 + \omegab^2{}_3 \dd\thetab^1 \\
  &= \left(2 f_0 f_1 \thetab^0 + (f_0^2 + f_1^2)
  \thetab^1 + 2f_0 \omegab^0{}_1\right) \thetab^2\thetab^3 -
  \omegab^2{}_3 \omegab^0{}_1 \thetab^0.
\end{aligned}
\end{equation}
Using \eqref{eq:dL0_appB} and \eqref{eq:S0_simple} we write the equation
$\mathbf dL_0=S_0$ (which is \eqref{eq:vacuum_equations_Einstein} for $i=0$)
in the form
\[
\left(2\dd f_1 + 2 f_1 \frac{\dd R^2}{R^2} - \frac1{R^2}
  \thetab^1 - 2 f_0 f_1 \thetab^0 - (f_0^2 + f_1^2)
  \thetab^1 - 2f_0 \omegab^0{}_1\right)\thetab^2\thetab^3=0
\]
and contract with two horizontal vectors. Then we find the equation
\begin{align}
\label{eq:dL0S0_simplified}
2\dd f_1 + 2 f_1 \frac{\dd R^2}{R^2} - \frac1{R^2}
  \thetab^1 - 2 f_0 f_1 \thetab^0 - (f_0^2 + f_1^2)
  \thetab^1 - 2f_0 \omegab^0{}_1 = 0.
\end{align}
Conversely, if \eqref{eq:dL0S0_simplified} holds then the field equation
$\dd L_0 = S_0$ is satisfied. In a similar way we can treat the
equation $\dd L_1 = S_1$, i.e. \eqref{eq:vacuum_equations_Einstein} for
$i=1$. We obtain
\begin{align}
\label{eq:dL1S1_simplified}
2\dd f_0 + 2 f_0 \frac{\dd R^2}{R^2} + \frac1{R^2}
  \thetab^0 - 2 f_0 f_1 \thetab^1 - (f_0^2 + f_1^2)
  \thetab^0 - 2f_1 \omegab^0{}_1 = 0.
\end{align}

For $i=2,3$ equation \eqref{eq:vacuum_equations_Einstein} has to be treated
differently. Let us consider the case $i=2$. We first define the invariant
1-form
\begin{align}
\label{eq:def_beta}
\betab = f_0\thetab^1 + f_1\thetab^0
\end{align}
and then we derive from \eqref{eq:L02_simple}
\begin{align}
\dd L_2 = -\dd (\betab + \omegab^0{}_1) \thetab^3 + (\betab +
\omegab^0{}_1) \dd\thetab^3 = -\alphab \omegab^0{}_1 \thetab^3 - (\betab +
\omegab^0{}_1) \omegab^3{}_2\thetab^2.
\end{align}
Using the fact that
\begin{align}
\dd\thetab^3 = - \omegab^3{}_0\thetab^0 - \omegab^3{}_1\thetab^1 -
\omegab^3{}_2\thetab^2 = - \omegab^3{}_2\thetab^2 + \alphab\thetab^3
\end{align}
we can write the equation $\mathbf dL_2=S_2$ as
\[
\left( \dd \betab + \dd \omegab^0{}_1 + \alphab\betab  \right) \thetab^3=0.
\]
Contraction with one horizontal vector shows that this equation
implies
\begin{align}
\label{eq:dL2S2_simplified}
\dd \betab + \dd \omegab^0{}_1 + \alphab\betab = 0
\end{align}
and, conversely, if \eqref{eq:dL2S2_simplified} is satisfied then also
$\dd L_2 =S_2$ holds. The equation $\dd L_3=S_3$ does not contribute anything
new.

Now we can collect the equations. The contents of the first structure
equation is the specific form of the connection forms, the first
structure equation on $S^2$ and the first structure equation on
$\cM_1$. Since all the relevant quantities are invariant they are
pull-backs from $\cM_1$. So the equations really live on
$\cM_1$. However, in order not to complicate the notation we will
continue to write them as equations on $\cM$, keeping in mind that
everything has to be regarded as a pull-back from the 2-dimensional
manifold $\cM_1$. Then we have
\begin{equation}
  \label{eq:1}
  \begin{aligned}
    \dd \thetab^0 + \omegab^0{}_1\thetab^1 &= 0\\
    \dd \thetab^1 + \omegab^1{}_0\thetab^0 &= 0.
  \end{aligned}
\end{equation}
Furthermore, we have the relationship between the 1-form $\alphab$ and
the differential of~$R^2$
\begin{equation}
  \label{eq:2}
  \dd R^2 = 2\alphab R^2.
\end{equation}
That means the integral of $\alphab$ along a curve is the same as the
difference of the values of $(\log R)$ at the boundary of that curve.
Hence $\alphab$ is related the velocity with that the area of the spheres
changes.

Then the field equations written in terms of $f_0$, $f_1$ and
$\betab=f_0\thetab^1+f_1\thetab^0$ are
\begin{align}
\label{eq:df0}
  \dd f_0 - \omegab f_1 + f_0 \alphab + \frac12\left(f_0^2 -
    f_1^2 + \frac1{R^2}\right)\thetab^0 &= 0,\\ 
\label{eq:df1}
  \dd f_1 - \omegab f_0 + f_1 \alphab - \frac12\left(f_0^2 -
    f_1^2 + \frac1{R^2}\right)\thetab^1 &= 0,\\
\label{eq:dbetapdomega}
  \dd \betab + \dd \omegab + \alphab\betab &= 0.
\end{align}
Note, that contracting the first of these equations with $\thetab^0$
and the second with $\thetab^1$ yields the integrability condition
$\dd\alphab=0$. On the other hand, using these two equations to
compute the differential of $\betab$ we find
\begin{equation}
  \label{eq:5}
  \dd\betab = -2 \alphab \betab - \frac1{R^2} \thetab^0\thetab^1
\end{equation}
and, therefore,
\begin{equation}
  \label{eq:6}
  \dd\omegab = \alphab \betab + \frac1{R^2} \thetab^0\thetab^1.
\end{equation}
Note, that $\alphab$ and $\betab$ are not independent. They contain
the same information. In fact, we have $\betab=\star \alphab$. This
relationship can be expressed in the present case also as two 2-form
equations
\begin{equation}
  \label{eq:7}
  \alphab\thetab^1 + \betab\thetab^0 = 0,\qquad
  \alphab\thetab^0 + \betab\thetab^1 = 0.
\end{equation}
This concludes the derivation of the reduced equations.

\end{document}

%% file: Stokes.pstex_t
\begin{picture}(0,0)%
\includegraphics{Stokes.pstex}%
\end{picture}%
\setlength{\unitlength}{4144sp}%
\begingroup\makeatletter\ifx\SetFigFont\undefined%
\gdef\SetFigFont#1#2#3#4#5{%
  \reset@font\fontsize{#1}{#2pt}%
  \fontfamily{#3}\fontseries{#4}\fontshape{#5}%
  \selectfont}%
\fi\endgroup%
\begin{picture}(2137,1563)(3646,-2074)
\put(3646,-2041){\makebox(0,0)[lb]{\smash{{\SetFigFont{9}{10.8}{\rmdefault}{\mddefault}{\updefault}{\color[rgb]{0,0,0}$n_0$}%
}}}}
\put(5446,-2041){\makebox(0,0)[lb]{\smash{{\SetFigFont{9}{10.8}{\familydefault}{\mddefault}{\updefault}{\color[rgb]{0,0,0}$n_1$}%
}}}}
\put(5131,-1276){\makebox(0,0)[lb]{\smash{{\SetFigFont{9}{10.8}{\familydefault}{\mddefault}{\updefault}$[n_1,n_2]$}}}}
\put(3736,-1186){\makebox(0,0)[lb]{\smash{{\SetFigFont{9}{10.8}{\familydefault}{\mddefault}{\updefault}$[n_0,n_2]$}}}}
\put(4636,-601){\makebox(0,0)[lb]{\smash{{\SetFigFont{9}{10.8}{\familydefault}{\mddefault}{\updefault}{\color[rgb]{0,0,0}$n_2$}%
}}}}
\put(4308,-1474){\makebox(0,0)[lb]{\smash{{\SetFigFont{9}{10.8}{\familydefault}{\mddefault}{\updefault}$[n_0,n_1,n_2]$}}}}
\put(4376,-1990){\makebox(0,0)[lb]{\smash{{\SetFigFont{9}{10.8}{\familydefault}{\mddefault}{\updefault}$[n_0,n_1]$}}}}
\end{picture}%

%% file: mesh.pstex_t
\begin{picture}(0,0)%
\includegraphics{mesh.pstex}%
\end{picture}%
\setlength{\unitlength}{4144sp}%
\begingroup\makeatletter\ifx\SetFigFont\undefined%
\gdef\SetFigFont#1#2#3#4#5{%
  \reset@font\fontsize{#1}{#2pt}%
  \fontfamily{#3}\fontseries{#4}\fontshape{#5}%
  \selectfont}%
\fi\endgroup%
\begin{picture}(4666,2086)(3871,-3035)
\put(4816,-2986){\makebox(0,0)[lb]{\smash{{\SetFigFont{12}{14.4}{\familydefault}{\mddefault}{\updefault}{\color[rgb]{0,0,0}initial simplicial complex $\mathcal C_i$}%
}}}}
\put(3871,-1756){\makebox(0,0)[lb]{\smash{{\SetFigFont{12}{14.4}{\familydefault}{\mddefault}{\updefault}{\color[rgb]{0,0,0}light-like}%
}}}}
\put(3931,-1126){\makebox(0,0)[lb]{\smash{{\SetFigFont{12}{14.4}{\familydefault}{\mddefault}{\updefault}{\color[rgb]{0,0,0}upwards-directed}%
}}}}
\put(7876,-2581){\makebox(0,0)[lb]{\smash{{\SetFigFont{12}{14.4}{\familydefault}{\mddefault}{\updefault}{\color[rgb]{0,0,0}time-step}%
}}}}
\put(6721,-1231){\makebox(0,0)[lb]{\smash{{\SetFigFont{12}{14.4}{\familydefault}{\mddefault}{\updefault}{\color[rgb]{0,0,0}downwards-directed}%
}}}}
\end{picture}%

%% file: spherical.bbl
\begin{thebibliography}{10}

\bibitem{regge1961:_gener_relat}
Regge T, 1961 {General Relativity} without coordinates.
\newblock \emph{Nouvo Cimento} \textbf{19} 558--571.

\bibitem{gentle2002:_regge}
Gentle A~P, 2002 Regge calculus: a unique tool for numerical relativity.
\newblock \emph{Gen. Rel. Grav.} \textbf{34} 1701--1718.

\bibitem{Thornburg:cqg21_3665}
Thornburg, J., 2004 Black-hole excision with multiple grid patches.
\newblock \emph{Class. Quant. Grav.} \textbf{21} 3665--3691

\bibitem{Caltech:gr-qc/0607056}
Scheel, M.~A. et~al., 2006
Solving Einstein's Equations With Dual Coordinate Frames.
\newblock gr-qc/0607056

\bibitem{LSU:cqg21_S553}
Schnetter, E. et~al., 2006
A multi-block infrastructure for three-dimensional time-dependent numerical
relativity
\newblock \emph{Class. Quant. Grav.} \textbf{23} S553 -- S578

\bibitem{frauendiener2006:_discr_differ_forms_gener_relat}
Frauendiener J, 2006 Discrete differential forms in {G}eneral {R}elativity.
\newblock \emph{Class. Quant. Grav.} \textbf{23} S369--S385.

\bibitem{bossavit1998:_discr_em_prob}
Bossavit A, 1998--2000 Discretization of electromagnetic problems.
\newblock \emph{Tech. rep.}, Interdyscyplinary Centre For Mathematical And
  Computational Modelling, Warsaw.
\newblock Http://www.icm.edu.pl/edukacja/mat/DEP.php.

\bibitem{bossavit1988:_mixed_whitn}
Bossavit A, 1988 Mixed finite elements and the complex of {W}hitney forms.
\newblock In \emph{The Mathematics of Finite Elements and Applications VI}, ed.
  J~Whiteman (London: Academic Press). pp. 137--144.

\bibitem{nedelec1980:_mixed_r}
N\'ed\'elec J~C, 1980 Mixed finite elements in {$\mathbb R^3$}.
\newblock \emph{Numer. Math.} \textbf{35} 315--341.

\bibitem{raviartthomas1977:_mixed_fem}
Raviart P~A and Thomas J~M, 1977 A mixed finite element method for 2nd order
  elliptic problems.
\newblock In \emph{Mathematical aspects of the finite element method}, eds.
  I~Galligani and E~Magenes (Berlin and New York: Springer-Verlag), vol. 606 of
  \emph{Lecture Notes in Mathematics}.

\bibitem{hiptmair2002:_finit_elem_disc_em}
Hiptmair R, 2002 Finite elements in computational electromagnetism.
\newblock \emph{Acta Numerica} \textbf{11} 237--339.

\bibitem{bossavit1998:_comput_elect}
Bossavit A, 1998 \emph{Computational Electromagnetism} (Boston: Academic
  Press).

\bibitem{PSU:PRD73_044028}
Sopuerta, C.~F. and Laguna, P., 2006 Finite element computation of the
gravitational radiation emitted by a pointlike object orbiting a
nonrotating black hole
\newblock \emph{Phys. Rev. D} \textbf{73} 044028

\bibitem{cartan2001:_rieman_geomet_in_orthog_frame}
Cartan {\'E}, 2001 \emph{Riemannian Geometry In An Orthogonal Frame}
  (Singapore: World Scientific).
\newblock Lectures delivered by E. Cartan at the Sorbonne 1926-27.

\bibitem{sparling2001:_twist_einst}
Sparling G, 2001 Twistors, spinors and the {Einstein} equations.
\newblock In \emph{Further advances in twistor theory {III}: {Curved} twistor
  spaces}, eds. L~J Mason, L~P Hughston, P~Z Kobak and K~Pulverer (Boca Raton:
  Chapman and Hall). pp. 179--187.

\bibitem{Wikipedia:Tetrad}
Landau, L D and Lifschitz, E F, 1975 \emph{The Classical Theory of Fields}
(Oxford: Pergamon Press)

\bibitem{frankel1997:_physic}
Frankel T, 1997 \emph{The geometry of Physics -- An introduction} (Cambridge:
  Cambridge University press).

\bibitem{whitney1957:_geomet_integ_theor}
Whitney H, 1957 \emph{Geometric Integration Theory} (Princeton: Princeton
  University Press).

\bibitem{gradinaruhiptmair1999:_whitn}
Gradinaru V and Hiptmair R, 1999 Whitney elements on pyramids.
\newblock \emph{ETNA} \textbf{8} 154--168.

%\bibitem{PhysRevD.50.R6033}
%Malec, E  and \'O Murchadha, N, 1994
%Optical scalars and singularity avoidance in spherical spacetimes.
%\newblock \emph{Phys. Rev. D} \textbf{50} R6033--R6036

\bibitem{munkres1993:_simpl_compl_simpl_maps}
Munkres J~R, 1993 Simplicial complexes and simplicial maps.
\newblock In \emph{Elements of Algebraic Topology} (Perseus Press), chap. 1.2.
  pp. 7--14.

\bibitem{powell1970}
Powell M~J~D, 1970 A hybrid method for nonlinear equations.
\newblock In \emph{Numerical Methods for Nonlinear Algebraic Equations}, ed.
  P~Rabinowitz (Gordon and Breach).

\bibitem{gsl2005:_refer_manual}
Galassi M et~al., 2005 \emph{GNU Scientific Library Reference Manual}.
\newblock \url{http://www.gnu.org/software/gsl/manual/gsl-ref_toc.html}.

\bibitem{PenroseRindler}
Penrose, R. and Rindler, W., 1986 \emph{Spinors and space--time : {S}pinor and
  twistor methods in space-time geometry}
({Cambridge: Cambridge University Press})

\bibitem{Weinberg}
Weinberg, S., 1972 \emph{Gravitation and Cosmology : Principles and
Applications of the General Theory of Relativity}
(New York: Wiley).

\bibitem{corlessgonnet1996:_lamber_w}
Corless R~M et~al., 1996 On the {L}ambert {W}-function.
\newblock \emph{Adv. Comput. Math.} \textbf{5} 329--359.

\end{thebibliography}
